\documentclass[12pt,a4paper]{article}
\usepackage{amssymb}
\begin{document}
\title{No-go theorem for the classical Maxwell-Lorentz electrodynamics in 
odd-dimensional worlds}
\author{Andrew E. Chubykalo${}^\dagger$, Augusto Espinoza${}^\dagger$, 
and B. P. Kosyakov${}^\ddagger$}
\maketitle
\begin{center}
{${}^\dagger$ 
Escuela de F${\acute\imath}$sica, 
Universidad Aut\'onoma de Zacatecas, 
Apartado Postal C-580 Zacatecas 98068, Zacatecas,  
Mexico\\
${}^\ddagger$ Russian Federal Nuclear Center, 
Sarov, 607190 Nizhnii Novgorod Region, Russia \\
{\bf E-mail:} 
${\rm achubykalo@yahoo.com.mx}$, 
${\rm drespinozag@yahoo.com.mx}$, 
${\rm kosyakov@vniief.ru}$} 
\end{center}
\begin{abstract}
\noindent
{If the conventional Maxwell--Lorentz formulation of classical electrodynamics 
is adopted in a flat spacetime of arbitrary odd dimension, then the 
retarded vector potential $A^\mu$ generated by a point charge turns out to be 
pure gauge, $A^\mu=\partial^\mu\chi$.
By Gauss' law, the charge shows up as zero.
The classical electromagnetic coupling is thus missing from odd-dimensional 
worlds.
If the action is augmented by the addition of the Chern--Simons term, then the 
classical interaction picture in the three-dimensional world becomes nontrivial.}
\end{abstract}


\section{Introduction}
In this note we present a surprising result that the classical 
electromagnetic interaction, realized as the Maxwell--Lorentz theory,
is missing from odd-dimensional spacetimes. 
It is well known that the four-dimensional 
electrodynamics can be extended to any even dimensions to result in
a consistent theory \cite{k99}, \cite{k07}, \cite{k08}.
However, one can hardly conceive that a similar extension to 
odd-dimensional spacetimes gives rise to a classical picture in which any point
charge generates zero field strengths, and hence, by Gauss' law, the 
electromagnetic coupling is effectively vanishing.
On the other hand, nontrivial solutions to homogeneous Maxwell's equations 
are still available.
Therefore, only free classical Maxwellian fields may exist in such worlds.

The odd-dimensional physics is of concern to us not only for 
methodological reasons but also in relation to the holographic principle
(for a review see, e. g., \cite{Aharony},  \cite{Bousso}), 
whose rationale often leads one to consider systems living in an odd-dimensional 
bulk, as, say, in the ${\rm AdS_3/CFT_2}$ model, the popular setting for 
analysing the holographic correspondence.
The no-go theorem discussed here may be of utility in such studies.

Section 2 gives a proof of this theorem.
Notice that the no-go theorem is only valid for a genuine $(2n+1)$-dimensional 
realm, not for electromagnetic systems constrained in a $2n$-dimensional spacelike manifold
which is actually immersed in a higher-dimensional spacetime.
We take a closer look at this issue in the final section. 
In addition, we adduce an argument that the analytical form of interaction 
between the quantized Maxwell field and different charged fields is common to all dimensions. 
It transpires that if the action is augmented by the addition of the 
Chern--Simons term, then the three-dimensional classical picture becomes 
nontrivial.

\section{The Maxwell--Lorentz Theory in Odd-Dimensional Spacetimes}
We begin with the conventional formulation of the Maxwell--Lorentz 
electrodynamics in a flat $D$-dimensional world.
Let us set the metric $\eta_{\mu\nu}={\rm diag}(1,-1,\ldots,-1)$, and adopt 
units in which the speed of light equals unity.
We write the action 
\begin{equation}
{S}=
S_{\rm p}-\int d^{D}\!x\,\left(j^\mu A_\mu+ {\frac{1}{4\,\Omega_{D-2}}}\,
F_{\mu\nu} F^{\mu\nu}\right),  
\label
{S-em}
\end{equation}
where $S_{\rm p}$ is the mechanical part of the action responsible for the
particle behavior, $\Omega_{D-2}$ is the area of the $(D-2)$-dimensional 
unite sphere, 
$\Omega_{D-2}=2\,{\pi^{(D-1)/{2}}}/{\Gamma\left[({D-1})/{2}\right]}$,
and the field strength is expressed in terms of the vector potential, 
$F_{\mu\nu}=\partial_\mu A_\nu-\partial_\nu A_\mu$.
For $D=3$, the action (\ref{S-em}) should be augmented by the addition of the 
Chern--Simons term.
However, we would like to compare even- and odd-dimensional cases 
within a unified framework, so that we 
ignore for a while any augmentations and employ the action (\ref{S-em}),
which is well suited to the even-dimensional electrodynamics 
\cite{k99}, \cite{k07}, \cite{k08}.

Varying $A_{\mu }$ in (\ref{S-em}) gives Maxwell's equations: 
\begin{equation}
\partial _{\mu }F^{\mu \nu }=
\Omega _{D-2}j^{\nu }\,\,.
\label
{maxw}
\end{equation}
The linearity of Maxwell's equations makes it possible to restrict our 
consideration to the single-particle case.
A particle with a $\delta$-shaped distribution of the electric charge $e$,
moving along an arbitrary world line $z^\mu(s)$, gives rise to the 
current  
\begin{equation}
j^{\mu}(x)=e\int_{-\infty}^\infty\!ds\,v^{\mu}(s)\,
\delta ^{D}\biggl (x-z(s)\biggr)\,,
\label
{j-em}
\end{equation}
where $v^\mu =dz^\mu/ds ={\dot z}^\mu$ is the $D$-velocity of this particle.

If we impose the Lorenz condition on the vector potential to fix the gauge, 
$\partial_\mu A^\mu=0$, then (\ref{maxw}) becomes  
\begin{equation}
\Box A^\mu=\Omega_{D-2} j^\mu\,\,.
\label
{wave-equation-D}
\end{equation}                                          

The physically relevant solution to (\ref{wave-equation-D}) is 
\begin{equation}
A^\mu(x)=\int d^{D}\!x\, G_{\rm ret}(x-x')\,j^\mu(x')\,\,,
\label
{sol}
\end{equation}
where $G_{\rm ret}(x)$ is the retarded Green's function of the wave operator. 
Our main interest here is with odd-dimensions  $D=2n+3,\,n=0,1,\ldots$
In this case (with reference to, e.~g., \cite{Iwanenko}, \cite{k07}), 
we have
\begin
{equation}
G_{\rm ret}(x)=N_D^{-1}\,\theta(x_0)\left(\frac{d}{dx^2}\right)^n
\frac{\theta(x^2)}{\sqrt{x^2}}\,\,,
\quad
N_D
=\frac{\Gamma\left(n+\frac12\right)}{\sqrt{\pi}}\,\,.
\label
{Green-D}
\end{equation}
We insert (\ref{j-em}) and (\ref{Green-D}) in (\ref{sol}).
Denoting $R_\mu= x_\mu-z_\mu(s)$, where $x_\mu$ is the observation point, and 
$z_\mu(s)$ the emission 
point on the world line, we obtain
\begin{equation}
A^\mu(x)=\frac{e}{N_D}\int^{\infty}_{-\infty} ds\, v^\mu(s)\,\theta(R_0)
\left(\frac{d}{dR^2}\right)^n\frac{\theta(R^2)}{\sqrt{R^2}}\,\,.
\label
{sol2}
\end{equation}
Let us take the integration variable $\lambda=R^2$, and observe that 
$d\lambda/ds=-2R\cdot v$, to  yield
\begin{equation}
A^\mu(x)=
\frac{e}{2N_D}\int^{\infty}_{0} d\lambda\,\frac{v^\mu}{R\cdot v}
\left(\frac{d}{d\lambda}\right)^n\frac{\theta(\lambda)}{\sqrt{\lambda}}\,\,.
\label
{sol3}
\end{equation}
For $n\ge 1$, the integral (\ref{sol3}) diverges at $\lambda=0$.
This is due to the fact that the integrand of (\ref{sol}) is the product
of two singular distributions.
Since the retarded  Green's functions are singular in themselves, they are normally 
integrated with smooth sources $j^\mu(x)$. 
To assign a mathematical sense to the ill-defined expression (\ref{sol3}), 
we regularize it by the convention that the derivatives should act on the 
left and the surface term is ignored.
According to this regularization prescription, 
\begin{equation}
A^\mu(x)=
\frac{e(-1)^n}{2N_D}\int^{\infty}_0 \frac{d\lambda}{\sqrt{\lambda}}\, 
\left(\frac{d}{d\lambda}\right)^n\frac{v^\mu}{R\cdot v}
=
\frac{e(-1)^n}{2N_D}\int^{\infty}_0 \frac{d\lambda}{\sqrt{\lambda}}\, 
\frac{\partial}{\partial x_\mu} \left(\frac{d}{d\lambda}\right)^n
\log ({R\cdot v})\,\,.
\label
{sol4}
\end{equation}
With the estimations $R^\mu =O(\lambda^{1/2})$, $v^\mu=O(1)$ as 
$\lambda\to\infty$, valid for world lines that approach asymptotically to 
straight timelike lines, the integral (\ref{sol4}) converges uniformly 
for $n\ge 1$, that is, for  $D\ge 5$. 
Therefore, the order of integration and differentiation with respect to $x^\mu$
may be interchanged with the opposite one.
As to $n=0$, one may think of (\ref{sol4}) as 
\begin{equation}
A_\mu(x)=\frac{e}{2}\,\lim_{\epsilon\to 0} \,
\frac{\partial}{\partial x^\mu}
\int^{\infty}_0 \frac{d\lambda}{\sqrt{\lambda}}\, 
e^{-\epsilon\lambda}\,\log ({R\cdot v})\,\,.
\label
{sol5}
\end{equation}
Be it as it may, whenever  the  ${\partial}/{\partial x_\mu}$ and integration 
are commutative, we get $A^\mu=\partial^\mu\chi$. 

In fact, the interchangeability of these operations is a subtle issue.
The order of implementation of these operations must not be changed 
if the measure contains a singular component, such as the $\delta$-function or 
its derivatives, because the integrals are no longer uniformly convergent. 
To illustrate, we refer to $D=4$.
In lieu of (\ref{sol3}), we have 
\begin{equation}
A^\mu(x)=
e\int^{\infty}_{-\lambda_m} d\lambda\,\delta(\lambda)\,\frac{v^\mu}{R\cdot v}=
e\int^{\infty}_{-\lambda_m} d\lambda\,\delta(\lambda)\,
\frac{\partial}{\partial x^\mu} \log ({R\cdot v})\,\,,
\label
{sol-4D}
\end{equation}
where $\lambda_m$ is the maximal absolute value of the spacelike interval 
between a given point of observation $x^\mu$ and the world line.
The  ${\partial}^\mu$ must be applied prior to  
integration, and hence (\ref{sol-4D}) results in 
$e v^\mu/({R\cdot v})|_{\rm ret}$, the Li{\'e}nard--Wiechert
vector potential, rather than  
$e \partial^\mu [\log({R\cdot v})]_{\rm ret}$ which differs from the 
Li{\'e}nard--Wiechert vector potential in the term
$eR^\mu[(R\cdot{\dot v})-1]/(R\cdot v)|_{\rm ret}$.

If we integrate (\ref{maxw}) over a domain $V$ containing the charged
particle in the hyperplane perpendicular to the world line and take
into account the relation $F_{\mu\nu}=0$, we obtain
\begin{equation}
e\,{\Omega_{D-2}}=\oint_{V} d^{D-1}\!x\,\,\partial_\mu F^{\mu\nu} v_\nu=
\oint_{\partial{V}} d^{D-2}\!x\,\,n_\mu F^{\mu\nu} v_\nu=0\,\,.
\label
{Gauss-law}
\end{equation}

Of course, nontrivial solutions to equation (\ref{wave-equation-D}) with 
$j^\mu=0$ are still available.
Therefore, applying the principle of least action to (\ref{S-em}) and using the
retarded boundary condition gives the picture involving only free fields.
This completes the proof of the theorem which reads: 
The retarded interaction of the classical Maxwell field with point charges 
shows up as vanishing in odd-dimensional spacetimes.

Note that this result is insensitive to the form of the regularization 
prescription in the sense that all reasonable extensions of the definition 
of the ill-defined expression (\ref{sol3}) as a distribution lead inevitably 
to  $A^\mu=\partial^\mu\chi$.
By contrast, in the even-dimensional case,  no regularization will render 
the corresponding integral uniformly convergent; in fact, any
regularization is superfluous here, which is exemplified most clearly by 
(\ref{sol-4D}).

\section{Discussion and Outlook}
An apparent objection against the above result may sound as follows.
A straight uniformly charged infinite string generates a static two-dimensional 
field.
This setting provides a simple counterexample of the statement that a 
nontrivial electromagnetic field generated by a point source cannot exist in 
$(1+2)$-dimensional worlds.
However, this objection does not discriminate between a system living in a
genuine $(2n+1)$-dimensional realm and that constrained effectively 
in a $2n$-dimensional spacelike manifold (as with the static field due to the 
rectilinear charged string, which, owing to its cylindrical symmetry, is 
generally regarded as a `two-dimensional' field), but, in fact, immersed in a 
higher-dimensional spacetime. 

To appreciate the distinction between `genuine' and `effective', let us
compare the behavior of a charged particle in a genuine $(1+1)$-dimensional 
realm and that in the case that the particle is moving along a straight line 
in ambient space.
In the former case, the field strength contains only the electric component
$F_{01}$ but the magnetic field is absent which implies that the Poynting 
vector is identically zero,  so that the accelerated charged particle does not 
emit radiation \cite{k99}, \cite{k07}.
On the other hand, a charged particle moving along a straight line
in ordinary three-dimensional space emits radiation with the emission rate
proportional to the  square of its acceleration.
 
This consideration suggests that the physics of a genuine $(2n+1)$-dimensional 
realm may differ drastically from that of systems which are constrained in a 
$2n$-dimensional manifold.
Although such systems living in higher-dimensional ambient spaces can imitate 
some $(2n+1)$-dimensional pictures, their behavior need not be governed by the
laws of genuine $(2n+1)$-dimensional realms. 

It is interesting that the discussed no-go theorem is inherently classical.
Indeed, if we substitute the retardation condition for the 
St{\"u}ckelberg--Feynman boundary condition and carry out the Wick rotation in 
the complex energy plane, then the propagator of a free massless field in
quantum theory, defined in a flat $D$-dimensional spacetime, becomes
\begin{equation}
D_E(x)\propto\frac{1}{\left({x_E^2}\right)^{\frac{D}{2}-1}}\,\,,
\label
{q-propagator}
\end{equation}
where $x_E^2$ is the negatively defined Euclidean length squared of the radius 
vector $x^\mu$,
\begin{equation}
{x_E^2}=-(x_4^2+{\bf x}^2)\,\,, \quad x_4=-ix^0\,\,.
\label
{Euclidean-norm}
\end{equation}
Equation (\ref{q-propagator}) shows that the analytical form of the photon 
propagator, appearing in perturbative calculations, is common to both 
$D=2n+1$ and $D=2n$.
The basic features of quantum physics in the former case can be obtained from 
those in the latter case by the analytic continuation in (\ref{q-propagator}).    

It may be worth pointing out that Gauss' law holds true in any dimension 
because it results from the action principle applied to the action (\ref{S-em})
with arbitrary $D$, while the actual coupling of the electromagnetic
field and charged matter in the case $D=2n+1$ may occur both trivial and 
nontrivial according to which boundary condition is additionally imposed. 

We finally consider the effect of incorporation of the Chern--Simons term into
the $D=3$ electrodynamic action.
Let us proceed from  
the action \cite{DeserJackiwTempleton82}  
\begin{equation}
{S}=
S_{\rm p}-\int d^{3}\!x\,\left[j^\mu A_\mu+ {\frac{1}{8\pi}}\,
(F_{\mu\nu} F^{\mu\nu}-\mu\,\epsilon^{\alpha\beta\gamma}A_\alpha 
F_{\beta\gamma})\right].  
\label
{S-CS}
\end{equation}
The field equation which follows from (\ref{S-CS}) is
\begin{equation}
\partial_\alpha F^{\alpha\beta}+\mu\,{}^\ast\! F^\beta=2\pi j^\alpha \,\,,
\label
{S-CS-field-eq}
\end{equation}
where ${}^\ast\! F^\alpha={\scriptstyle\frac12}\,\epsilon^{\alpha\beta\gamma}
F_{\beta\gamma}$. 
We rewrite this equation as 
\begin{equation}
\Lambda^{\alpha\beta}(\partial)\,{}^\ast\! F_\beta =(\mu\,\eta^{\alpha\beta}+
\epsilon^{\alpha\beta\gamma}\partial_\gamma)\,{}^\ast\! F_\beta=2\pi j^\alpha\,\,,
\label
{S-CS-eq}
\end{equation}
iterate it with $\Lambda(\partial)$, and use the Bianchi identity 
$\partial_\beta{}^\ast\! F^\beta=0$, to obtain
\begin{equation}
(\Box+\mu)\,{}^\ast\! F_\alpha=2\pi \Lambda_{\alpha\beta}(\partial) j^\beta\,\,.
\label
{field-eq-CS}
\end{equation}

It is clear from (\ref{field-eq-CS}), even without explicit writing its 
solution, that the field strength  generated by a point charged particle is 
nonzero because equation (\ref{field-eq-CS}) allows expressing the field 
$F_{\alpha\beta}$ directly in terms of the source $j^\alpha$.
We thus see that the classical dynamics governed by the action (\ref{S-CS}) is 
nontrivial.

The initial version of this paper ``On classical electrodynamics in 
odd-dimensional worlds'', hep-th/0208170, by B. P. Kosyakov, was 
rewritten by the present authors. 
The main trouble with the previous version was that the author cannot parry 
the objection against the no-go theorem discussed in Section 3.
The three present authors together managed to meet the challenge.

\end{document}